\documentclass[a4paper]{jpconf}
\usepackage{graphicx}
\usepackage{wrapfloat}
\usepackage{cite}
\def\Vec#1{\mbox{\boldmath $#1$}}
      
\def\Ms{M_{\rm s}}      
\def\Jm{J_{\rm m}}
\def\cH{{\mathcal H}}
\def\vS{{\Vec S}}

\begin{document}
\title{Effect of monomer-monomer interactions on the phase diagrams of the $\mathbf{S=1/2}$ distorted diamond type quantum spin chain}

\author{Kiyomi Okamoto$^{*1}$, Takashi Tonegawa$^{\dagger 2,3}$ and T\^oru Sakai$^{\ddagger 4,5}$}

\address{$^1$School of Arts and Sciences, College of Engineering, Shibaura Institute of Technology, \\
Minuma-ku, Saitama-shi, Saitama, 337-8570, Japan \\
$^2$Professor Emeritus, Kobe University, Nada-ku, Kobe-shi, Hyogo,  657-8501, Japan \\
$^3$Department of Physical Science, School of Science, Osaka Prefecture University, Naka-ku, Sakai-shi, 
Osaka, 599-8531, Japan \\
$^4$Graduate School of Material Science, University of Hyogo, Kamigori-cho, Ako-gun, Hyogo,\\
678-1297, Japan \\
$^5$Japan Atomic Energy Agency, SPring-8, Sayo-cho, Sayo-gun, Hyogo, 679-5148, Japan }

\ead{$^*$nnn2411@yahoo.co.jp, $^\dagger$tone0115@vivid.ocn.ne.jp, $^\ddagger$sakai@spring8.or.jp}

\begin{abstract}
By use of mainly the exact diagonalization and the level spectroscopy method,
we investigate the ground-state phase diagrams of the $S=1/2$ distorted diamond type quantum spin chain
with the monomer-monomer interactions and/or ferromagnetic interactions
for the zero magnetic field case, as well as the $M=\Ms/3$ case and the $M=(2/3)\Ms$ case,
where $M$ is the total magnetization and $\Ms$ is the saturation magnetization.
The magnetization plateau at $M=\Ms/3$ vanishes in the region
where the ferromagnetic interaction is rather strong.
The monomer-monomer interaction remarkably stabilizes the magnetization plateau at
$M=(2/3)\Ms$.

\end{abstract}

\section{Introduction}

In recent years,
the frustration effects on low-dimensional quantum spin systems have been 
attracting increasing attention.
The $S=1/2$ distorted diamond type quantum spin chain \cite{okamoto1},
sketched in Fig.1.
is known as an interesting model due to
the frustration, the quantum nature of spins, the low dimensionality and the trimer nature.
The Hamiltonian for this model is
described by 
\begin{eqnarray}
    \cH
    &=&   J_1 \sum_j \left( \vS_{3j-1} \cdot \vS_{3j} + \vS_{3j} \cdot \vS_{3j+1} \right)
        + J_2 \sum_j \vS_{3j+1} \cdot \vS_{3j+2} \nonumber \\
      &~& + \,J_3 \sum_j \left( \vS_{3j-2} \cdot \vS_{3j} + \vS_{3j} \cdot \vS_{3j+2} \right)
        + \Jm \sum_j \vS_{3j} \cdot \vS_{3(j+1)}
        - H \sum_j S_j^z
\end{eqnarray}
where $\vS_i = (S_i^x, S_i^y,S_i^z)$ is the spin-1/2 operator at the $i$th site,
and $H$ denotes the magnetic filed along the $z$ direction.
Hereafter we often use the normalized coupling constants
$\tilde J_i \equiv J_i/J_1$ for $i=1,2,3,{\rm m}$.
In the original version of this model \cite{okamoto1}, 
it was supposed that all of $J_1$, $J_2$ and $J_3$ are antiferromagnetic
and $\Jm=0$.
\begin{figure}[ht]
       \centerline{\scalebox{0.6}{\includegraphics{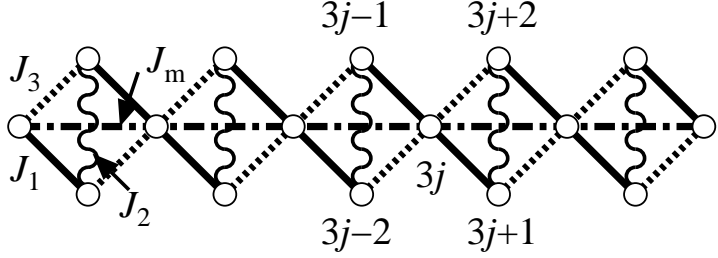}}}
       \label{fig:model-jm}
       \caption{The distorted diamond type quantum spin chain.
       Open circles denote $S=1/2$ spins, and various types of lines the interactions between spins.
       }
\end{figure}

Okamoto {\it et al.} \cite{okamoto1,tonegawa1,okamoto2}, and Honecker and L\"auchli \cite{honecker1}
discussed the ground-state properties for the above case.
The ground-state phase diagrams for the zero magnetic field case, the $M=\Ms/3$ case and the $M=(2/3)\Ms$ case
were obtained by Okamoto {\it et al.} \cite{okamoto1,tonegawa1,okamoto2},
where $M$ is the magnetization defined by $M \equiv \sum_j S_i^z$ which commutes with the total 
Hamiltonian $\cH$, and $\Ms$ is the saturation magnetization.
The zero magnetic field phase diagram consists of three phases,
the Tomonaga-Luttinger liquid (TLL) phase, the dimerized phase and the ferrimagnetic phase.
For the $\Ms/3$ case,
the magnetization plateau always exist except for the plateau mechanism changing line.
The $(2/3)\Ms$ magnetization plateau exists in a narrow region.

After the above pioneering works,
Kikuchi {\it el al.} (including present three authors) \cite{kikuchi1,kikuchi2}
reported that the magnetic properties of natural mineral azurite ${\rm Cu_3(CO_3)_2(OH)_2}$
can be well described by the original distorted diamond type chain mode, and estimated $J_1:J_2:J_3 = 1:1.25:0.45$.
Gu and Su \cite{gu1,gu2} stated that the double peak behavior of the susceptibility $\chi(T)$ can be
well fitted by the parameter set $J_1 : J_2 : J_{3z} = 1 : 1.9 : -0.3$
and $J_{3x}/J_{3z} = J_{3y}/J_{3z} = 1.7$.
Namely, Gu and Su claimed that the coupling $J_3$ is ferromagnetic and rather anisotropic,
while other couplings are antiferromagnetic and isotropic.
We did not agree their parameter set \cite{kikuchi3} because the angles of the $\rm{Cu-O-Cu}$
super exchange paths of $J_1$ and $J_3$ are very similar to each other
($113.7^\circ$ and $113.5^\circ$, respectively),
which means that $J_1$ and $J_3$ cannot differ in sign.
After the papers of Gu and Su,
some papers \cite{mikeska,kang} support $J_3>0$ (antiferromagnetic)
whereas others \cite{li,rule} $J_3<0$ (ferromagnetic).

This controversy was resolved by Jeschke {\it et al.} \cite{jeschke}
who pointed out the importance of the monomer-monomer interaction $\Jm$
by the first-principle calculation.
They showed that the magnetization curve $M(H)$ and the susceptibility $\chi(T)$
can be well fitted by $J_1:J_2:J_3:\Jm = 1:2.13:0.45:0.30$.
Honecker {\it et al.} \cite{honecker2} showed that the inelastic neutron scattering result \cite{rule} was
well explained by the above parameter set including $\Jm$.
However, the phase diagrams in the presence of $\Jm$ have not been discussed until now.

Very recently, experimental results on new distorted diamond type chain materials have been reported,
which are summarized in Table \ref{table:new}.
In some cases
the existences of $\Jm$ and ferromagnetic interactions are expected.
The nonexistence of the $Ms/3$ plateau in (d) is also important,
because the unit cell of the present model consists of three $S=1/2$ spins,
which often brings about the $\Ms/3$ plateau by the condition of Oshikawa, Yamanaka and Affleck \cite{oshikawa}.
As already stated, we \cite{okamoto1,tonegawa1,okamoto2}
obtained the phase diagrams of the present model for $J_1,J_2,J_3 \ge 0$ (antiferromagnetic) and $\Jm=0$ case.
Considering the above situations,
the phase diagrams with $\Jm$ and/or ferromagnetic interactions are urgently necessary.

In this paper,
we investigate the phase diagram of the present model with $\Jm$ and/or
ferromagnetic interactions by use of the exact diagonalization (ED) and 
the level spectroscopy (LS) method \cite{ls1,ls2,ls3,ls4}.
The phase diagrams are presented in \S2, and the discussion is given in \S3.

\begin{table}[h]
\label{table:new}
\caption{Summary of experimental results on azurite (a) and new distorted diamond type chain materials (b)-(g).
The materials for (c) is
${\rm \alpha- [Cu_3(OH)_2(CH_3CO_2)_2(H_2O)_4](C_6H_5SO_3)_2}$ and
that for (f) is ${\rm [Cu_3(OH)_2(CH_3CO_2)_2(OH)_2(H_2O)_2](2, 6 - Np)}$ where
Np is ${\rm C_{10}H_8(SO_3)_2}$.
In (b) ``A" denotes adipic acid ${\rm HOOC (CH_2)_4 COOH}$.
In (g), the combinations of ``A" and ``M" are
${\rm [A,M] = [K, Ga],~ [Rb, Al],~ [Rb, Ga],~ [Cs, Al]}$ and $\rm{[Cs, Ga]}$.
All of these materials except for (f) show gapless excitation behaviors at low temperatures,
which strongly suggests the TLL ground state. 
The spin gap behavior was found in (f),
which seems to be due to the dimerization 
(namely, the fact that the unit cell consists of six $S=1/2$ spins) coming from the crystal structure.}

\begin{center}
\begin{tabular}{lllllll}
\br
&ref.
&material(s) 
&$\Jm$ &\shortstack{ferromagnetic \\ interactions}
&\shortstack{$\Ms/3$ \\ plateau} &\shortstack{$(2/3)\Ms$ \\ plateau} \\
\mr
(a) 
&\cite{kikuchi1,kikuchi2}
&${\rm Cu_3(CO_3)_2(OH)_2}$ &yes &no &yes (wide) &no \\
\mr
(b) 
&\cite{kikuchi4,kikuchi5}
&$\rm{Cu_3(A)_2(OH)_2(H_2O)_4}$ &? &maybe &yes (narrow) &no \\
\mr
(c)
&\cite{yoneyama}
&see caption
&? &maybe &? &? \\
\mr
(d) 
&\cite{fujihala,koorikawa}
&${\rm K_3Cu_3AlO_2(SO_4)_4}$
&maybe &no &no &? \\
\mr
(e)
&\cite{asano}
&${\rm Cu_3(CrO_4)_2(OH)_2(C_5H_5N)_2}$
&? &maybe &yes (wide) &? \\
\mr
(f)
&\cite{fujita}
&see caption    
&? &maybe &? &? \\
\mr
(g)
&\cite{koorikawa}
&${\rm A_3Cu_3MO_2(SO_4)_4}$
&maybe &no &? &? \\
\br
&\end{tabular}
\end{center}
\end{table}

\section{Ground-state phase diagram}

To determine the phase diagrams,
we numerically performed the ED for finite systems up to 30 spins
by use of the Lancz\"os algorithm under the periodic and twisted boundary conditions. 

\subsection{zero magnetic field case}

It is very easy to distinguish the ferromagnetic ground-state ($M=\Ms$)
and the ferrimagnetic ground-state  state ($M = Ms/3$) 
from the $M=0$ ground-state.
It is expected that there are two phases, the TLL phase and the dimerized phase,
for the $M=0$ case.
The former has the gapless excitation and the latter the gapped excitation.
The phase boundary between these two phases is of the Berezinskii-Kosterlitz-Thouless (BKT) type,
which is rather difficult to numerically detect from the ED data.
We use the LS method \cite{ls1,ls2} which is a very powerful tool
to detect the BKT transition point from the ED data.
The details of the LS method for the present case are fully explained in \cite{okamoto1}.

Figure 2 shows the phase diagrams with $\tilde \Jm = 0$ and $\tilde \Jm=0.3$
at zero magnetic field.
The shape of the phase diagram with  $\tilde \Jm=0.3$ was slightly modified
from that with  $\tilde \Jm=0.0$.

\begin{figure}[ht]
       \centerline{
         \scalebox{0.3}{\includegraphics{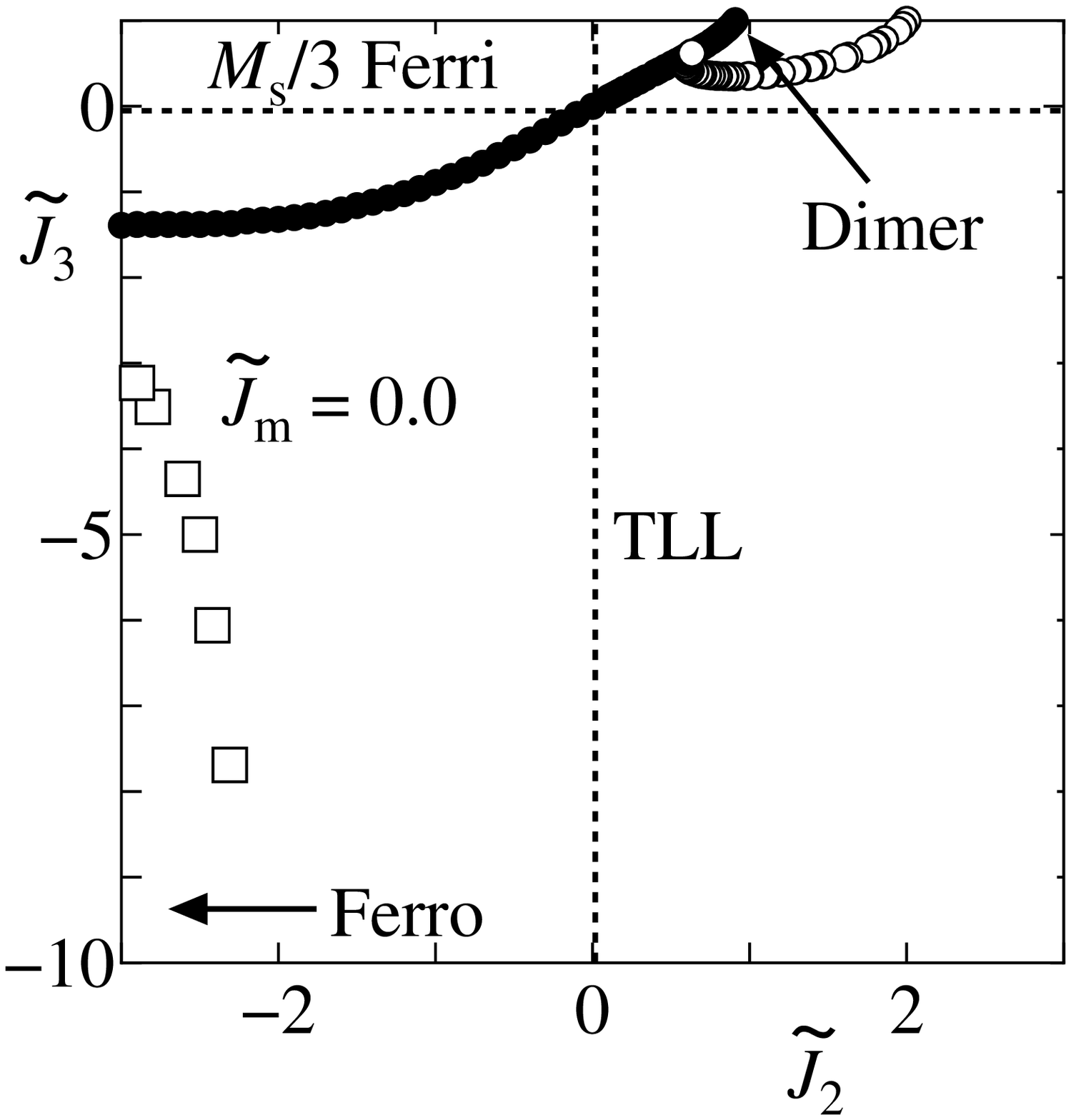}}~~~~
         \scalebox{0.3}{\includegraphics{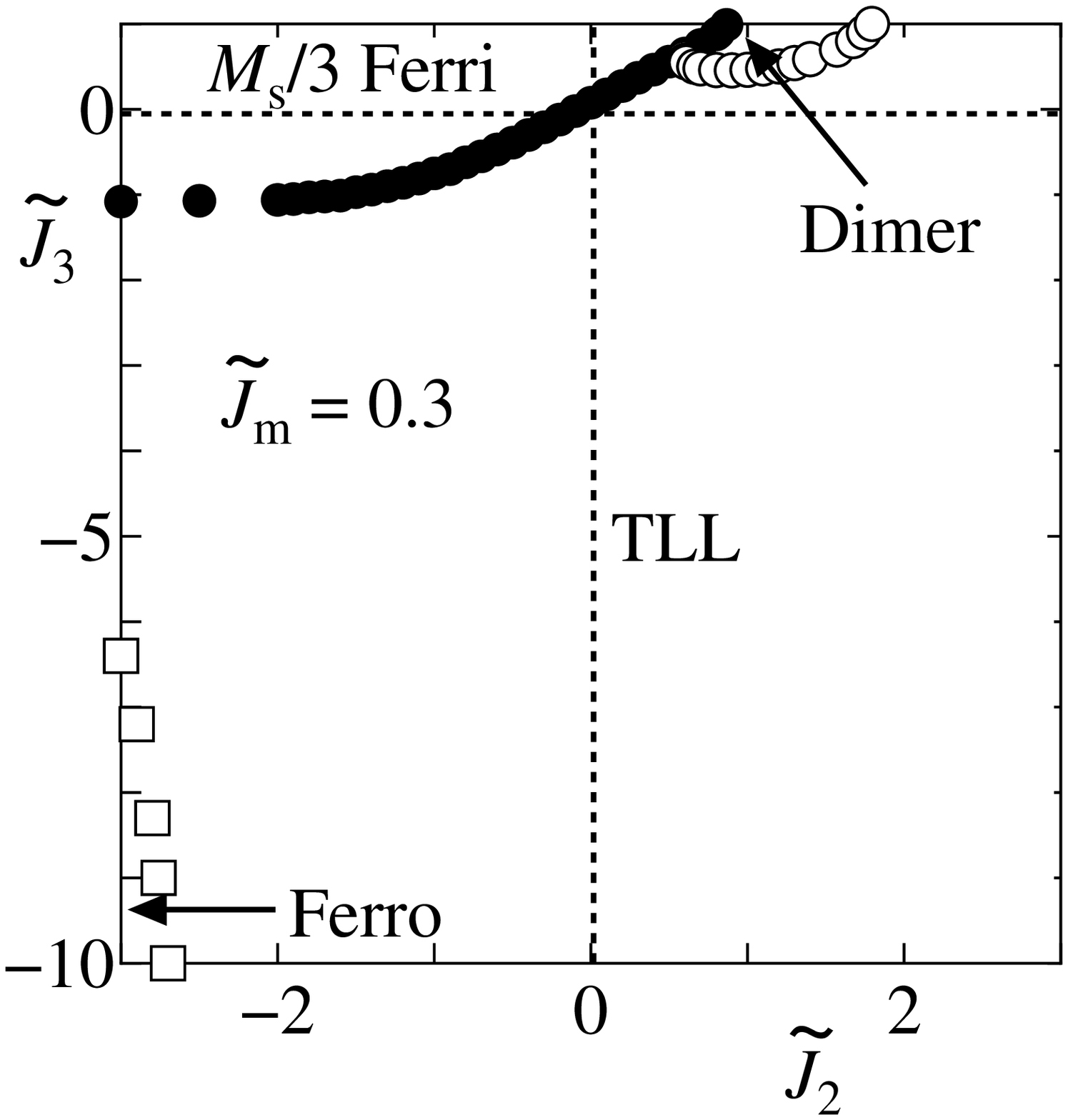}}~~~~
         \scalebox{0.3}{\includegraphics{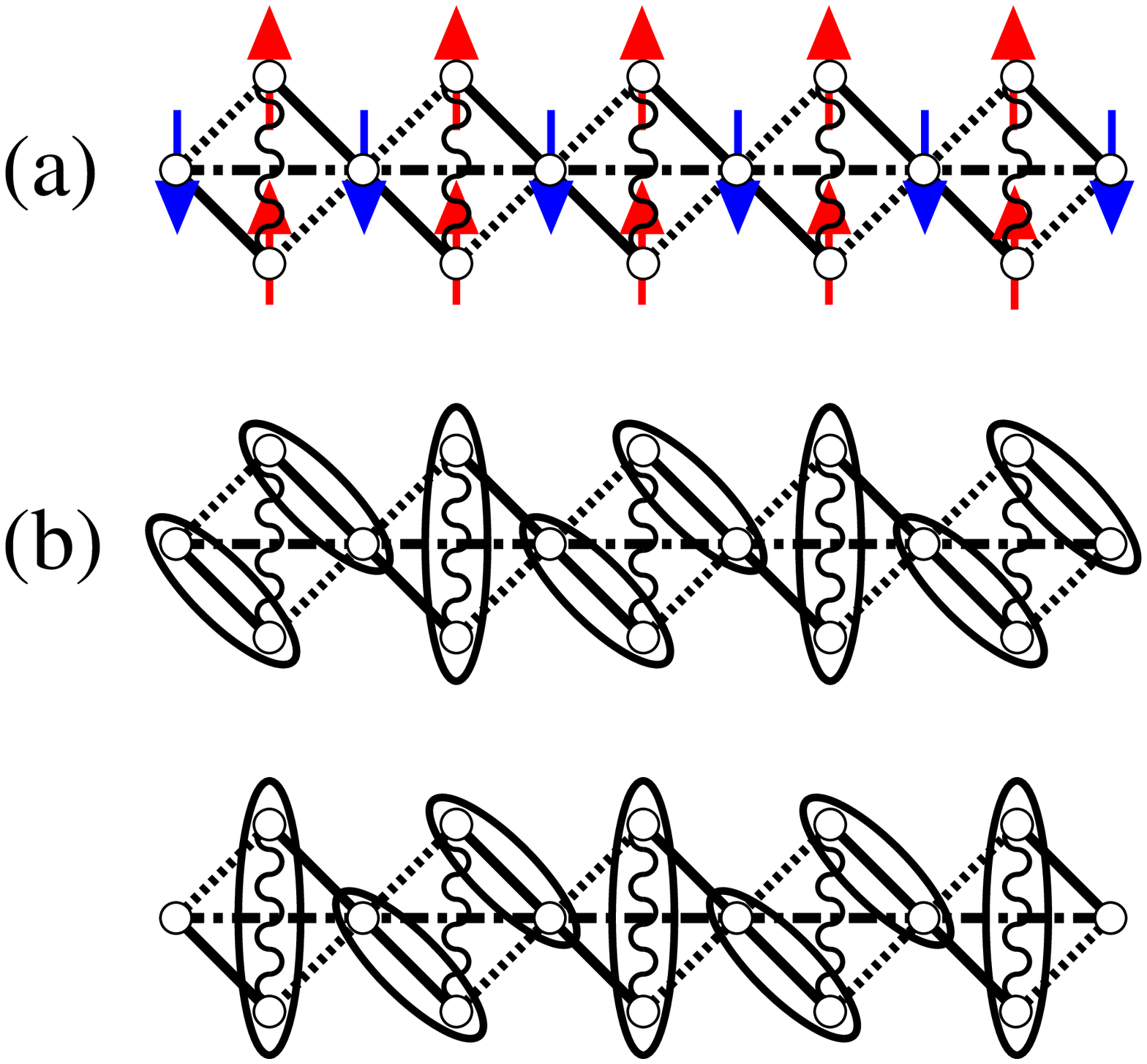}}
       }
       \label{fig:pd-m=0}
       \caption{The left and center panels are the ground-state phase diagrams
       for the zero magnetic field case with $\tilde \Jm = 0.0$ and $\tilde \Jm = 0.3$,
       respectively.
       The TLL-dimer boundary (\opencircle) is of the BKT type,
       while other boundaries (\opensquare \ and \fullcircle) are of the first order.
       The right panel shows the spin configurations for (a) the ferrimagnetic state
       and (b) the dimerized state, where ellipses denote the singlet pairs.
       The dimerized state is doubly degenerate.
       }
\end{figure}

\subsection{$M=\Ms/3$ case}

The detailed explanations of the LS method for the $M=\Ms/3$ case were given in \cite{okamoto2}.
The phase diagrams for the $M=\Ms/3$ case with $\tilde \Jm = 0.0$ and $\tilde \Jm = 0.3$,
as well as the spin configurations of the plateau states,
are given in Fig.2.
The appearance of the no $\Ms/3$ plateau region in the lower-left part
should be noticed.
The addition of $\tilde \Jm = 0.3$ slightly modifies the whole shape of the phase diagram.

\begin{figure}[ht]
       \centerline{
         \scalebox{0.3}{\includegraphics{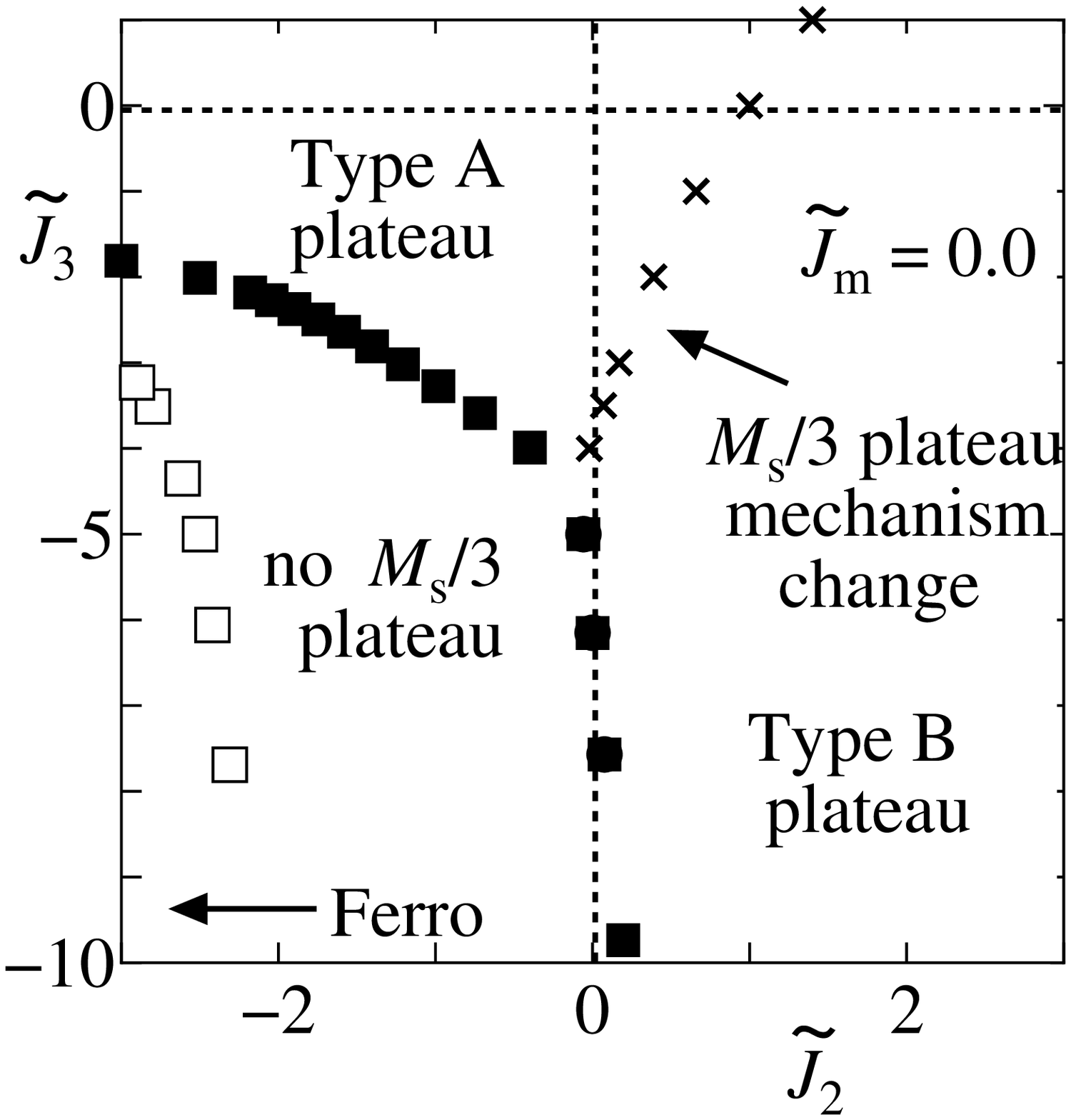}}~~~~
         \scalebox{0.3}{\includegraphics{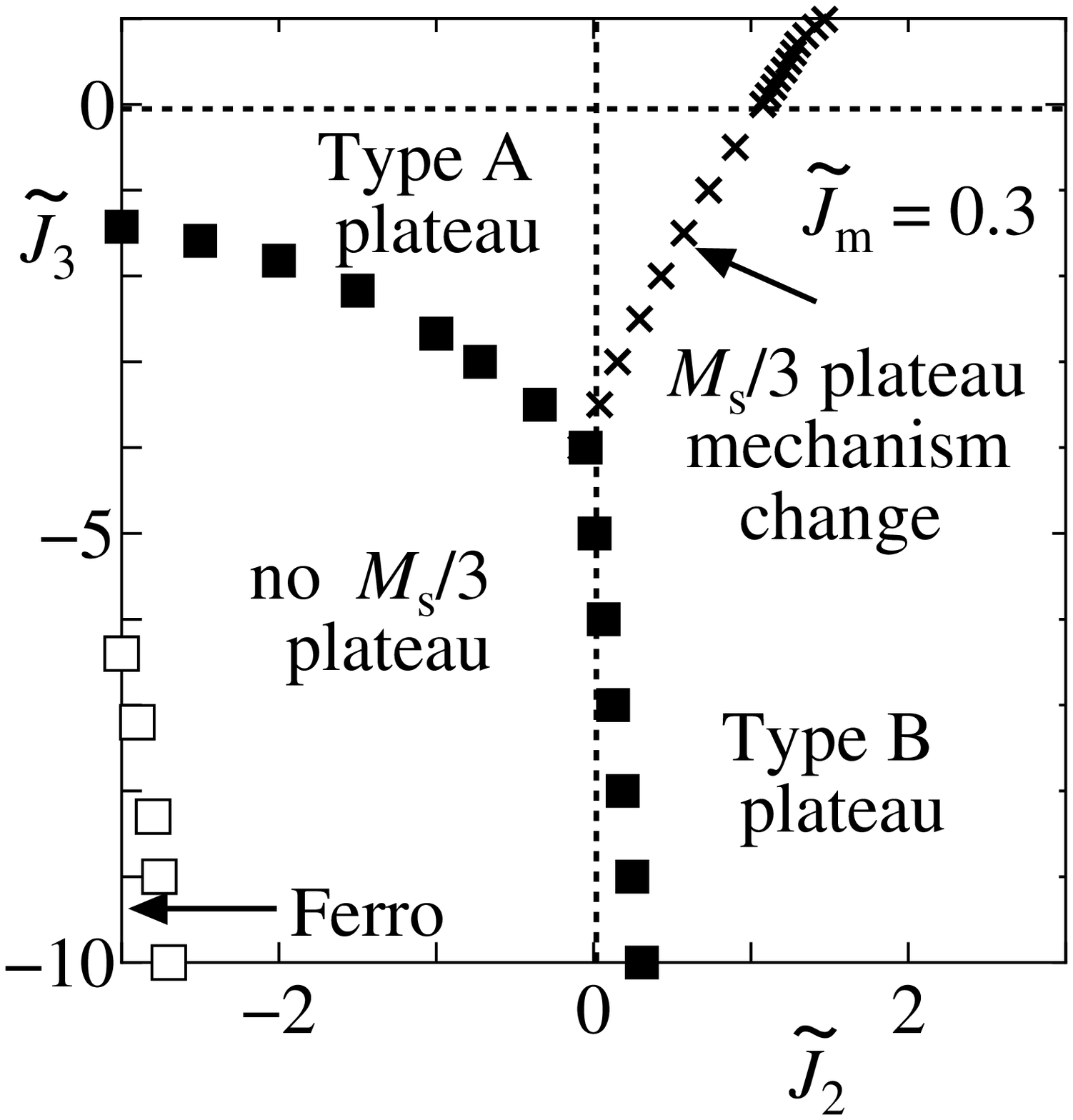}}~~~~
         \scalebox{0.3}{\includegraphics{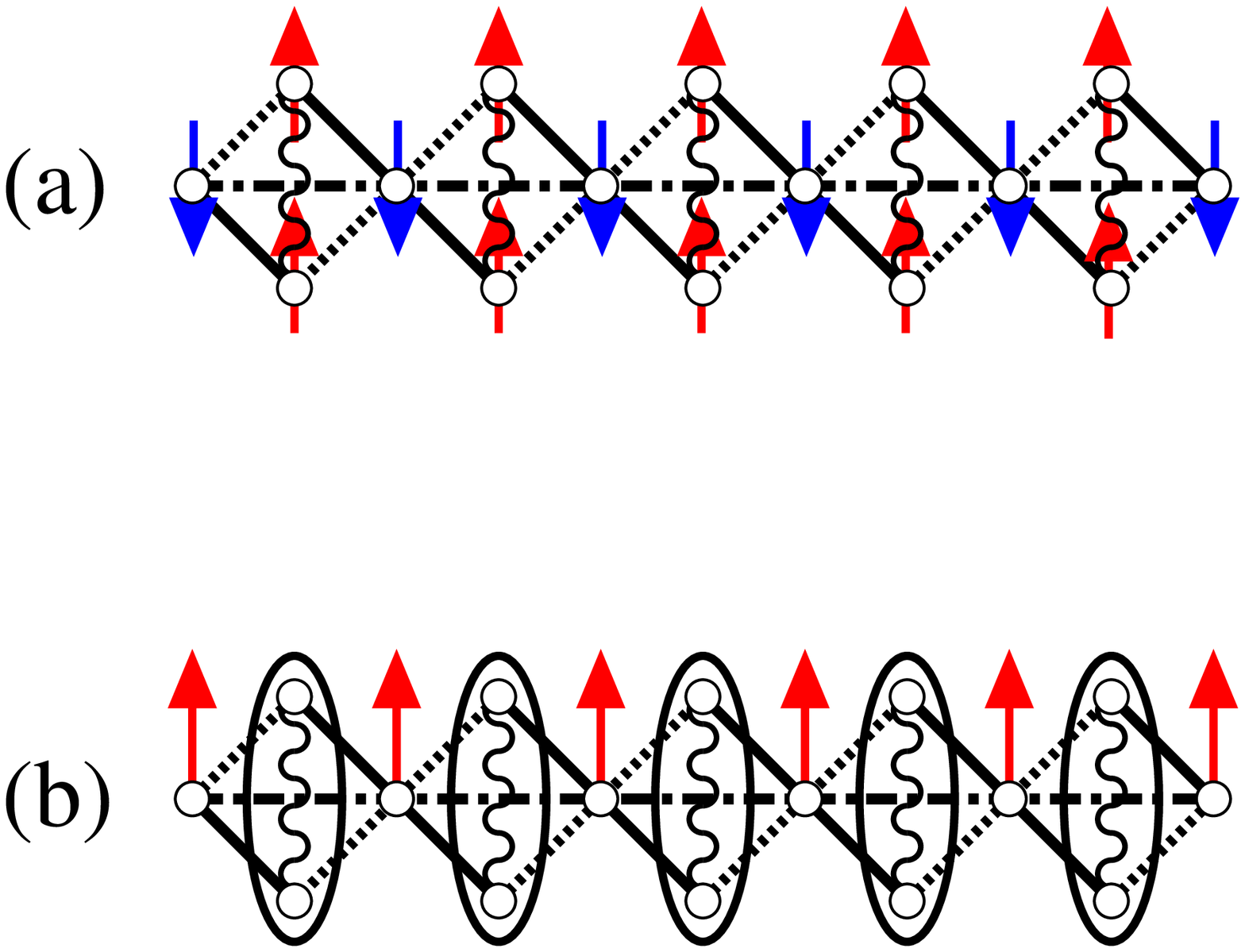}}
       }
       \label{fig:pd-m=1-3}
       \caption{The left and center panels are the ground-state phase diagrams
       for the $M=\Ms/3$ case with $\tilde \Jm = 0.0$ and $\tilde \Jm = 0.3$,
       respectively.
       The boundary (\opensquare) between the ferromagnetic phase and the no $\Ms/3$ plateau phase
        is of the first order,
       and that (\fullsquare) between the $\Ms/3$ plateau phase to the no $\Ms/3$ plateau phase 
       is of the BKT type.
       The plateau mechanism changing line ($\times$) is of the Gaussian type.
       The right panel shows the spin configurations for (a) the type A plateau and
       (b) the type B plateau,  
       where ellipses denote the singlet pairs.
       The spin configuration of the type A plateau is the same as that of the
       ferrimagnetic state under zero magnetic field in the right panel of Fig.2.
       }
\end{figure}

\subsection{$M=(2/3)\Ms$ case}

Since the unit cell of the present model consists of three $S=1/2$ spins,
the $M=(2/3)\Ms$ magnetization plateau is expected associated with spontaneous symmetry breaking
of the translational invariance \cite{oshikawa}.
The LS method for the $M=(2/3)\Ms$ case was explained in \cite{okamoto2}.
Figure 4 shows the phase diagrams for the $M=(2/3)\Ms$ case with
$\tilde \Jm = 0.0$, $\Jm = 0.1$ and $\tilde \Jm = 0.3$.
We can see that the addition of $\tilde \Jm$ drastically extends the plateau region
towards the upper left direction.

\begin{figure}[ht]
       \centerline{
         \scalebox{0.3}{\includegraphics{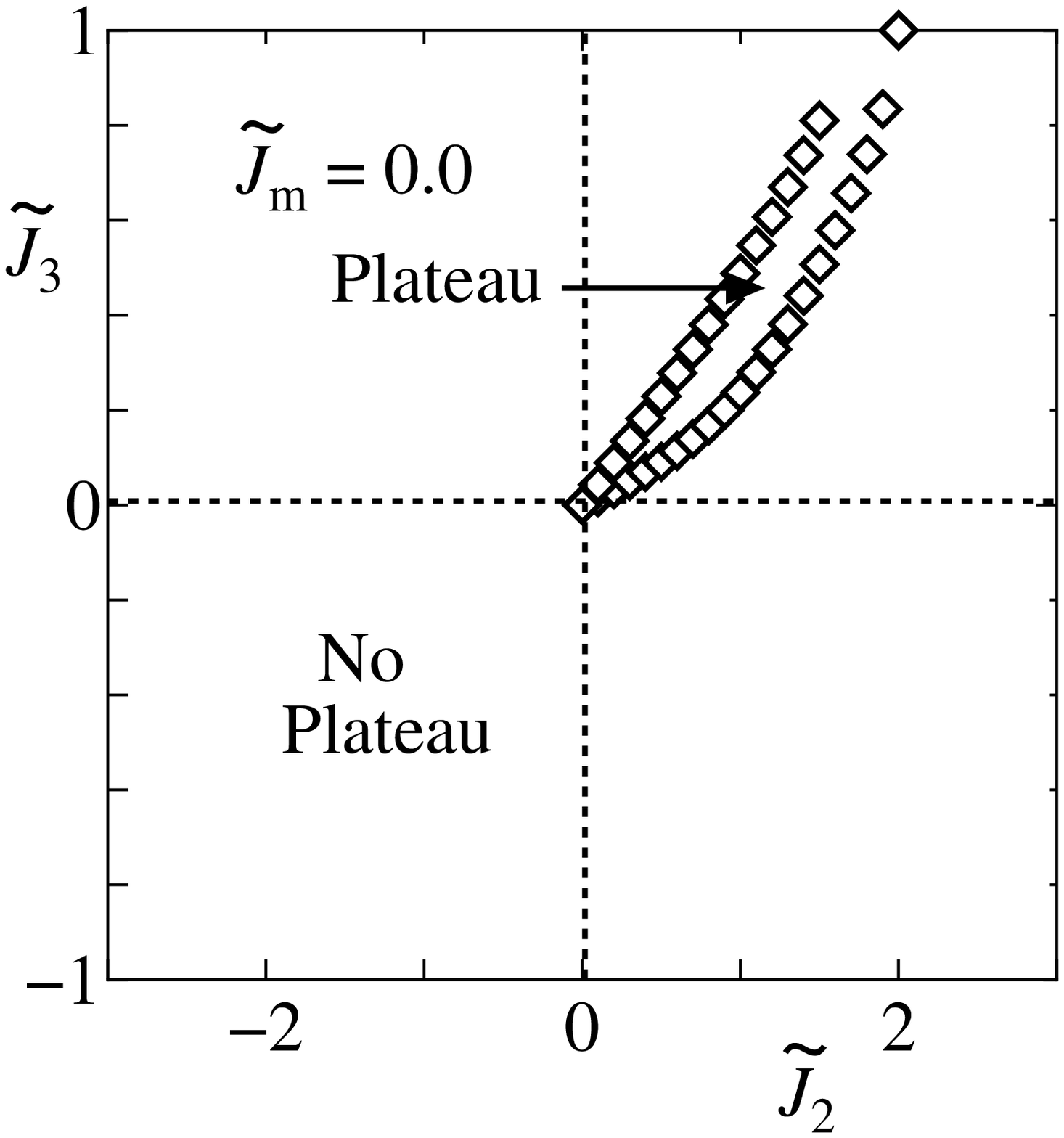}}~~~~~~
         \scalebox{0.3}{\includegraphics{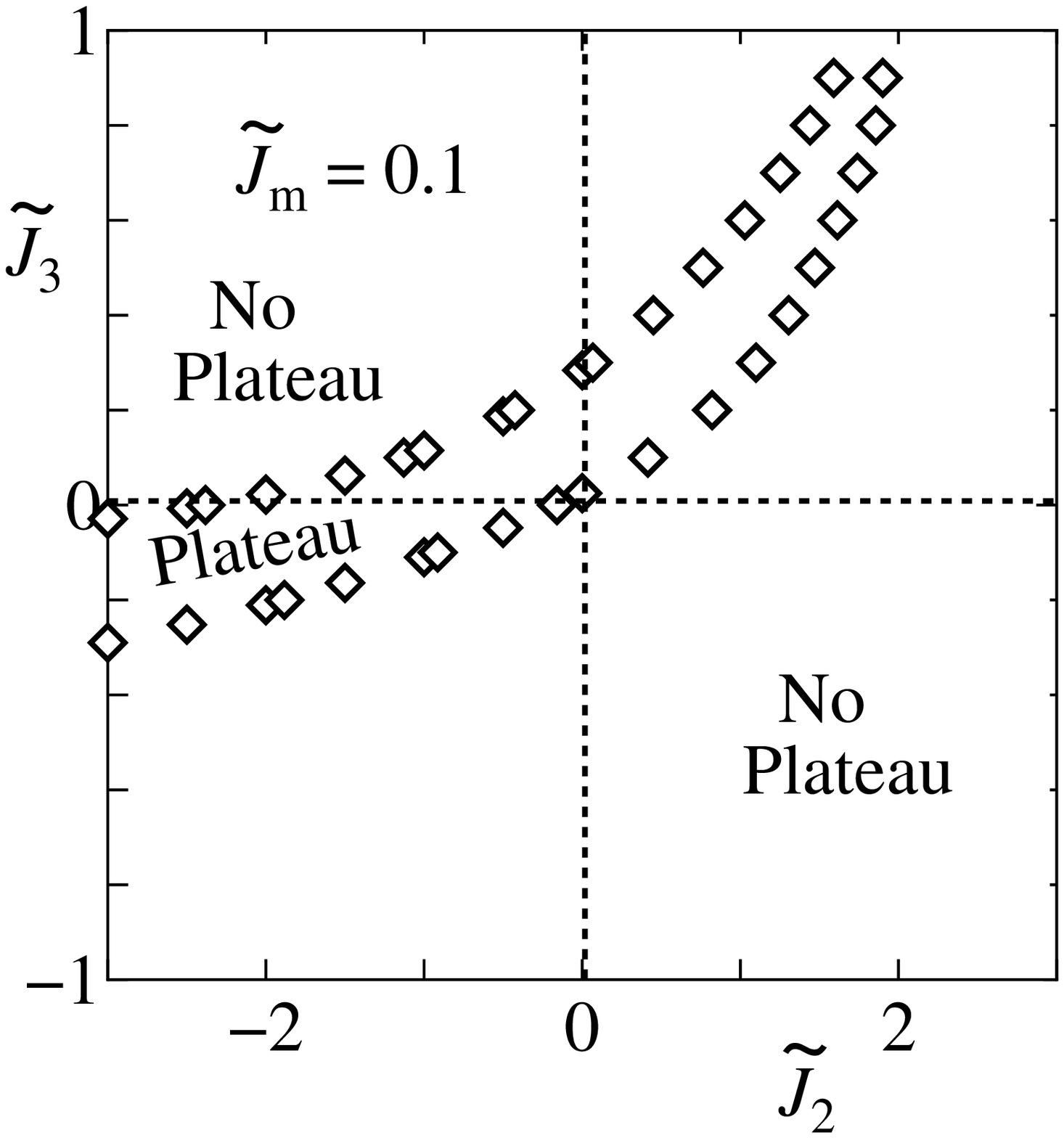}}~~~~~~
         \scalebox{0.3}{\includegraphics{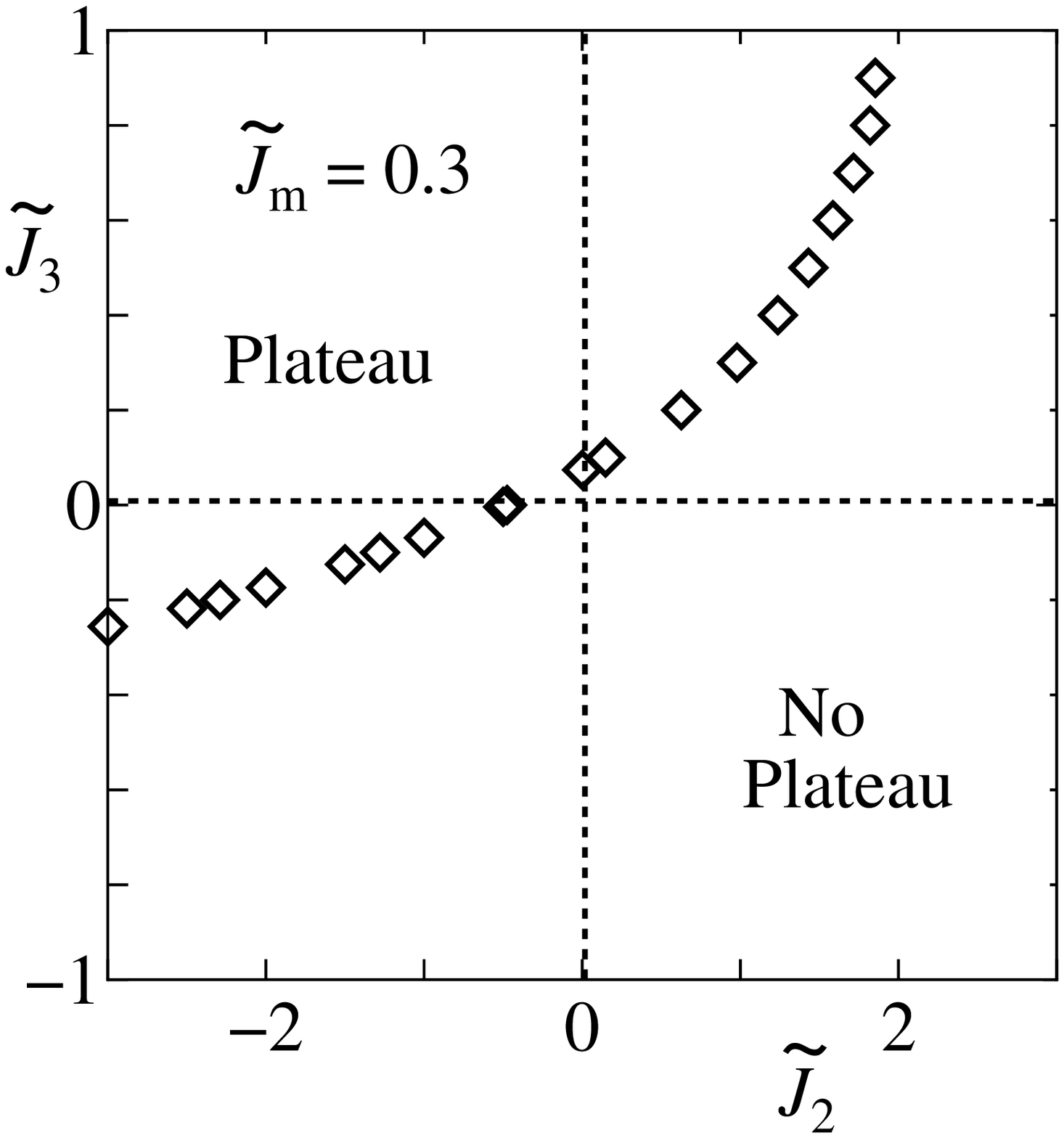}}
       }
       \label{fig:pd-m=2-3}
       \caption{Ground-state phase diagrams for the $M=(2/3)\Ms$ case
        with $\tilde \Jm = 0.0$ (left), $\tilde \Jm = 0.1$ (center) and $\tilde \Jm = 0.3$ (right).
       The boundaries (\opendiamond) between the plateau phase and the no plateau phase 
       are of the BKT type.
       }
\end{figure}

\begin{figure}[ht]
       \centerline{
       \scalebox{0.3}{\includegraphics{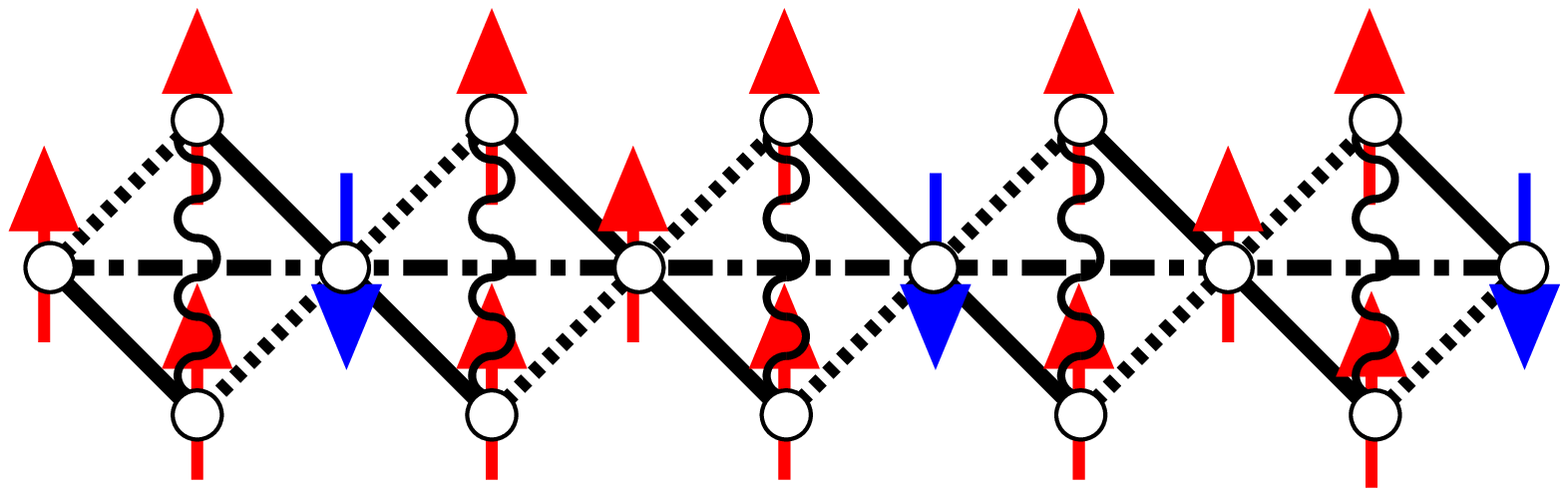}}~~~~~
       \scalebox{0.3}{\includegraphics{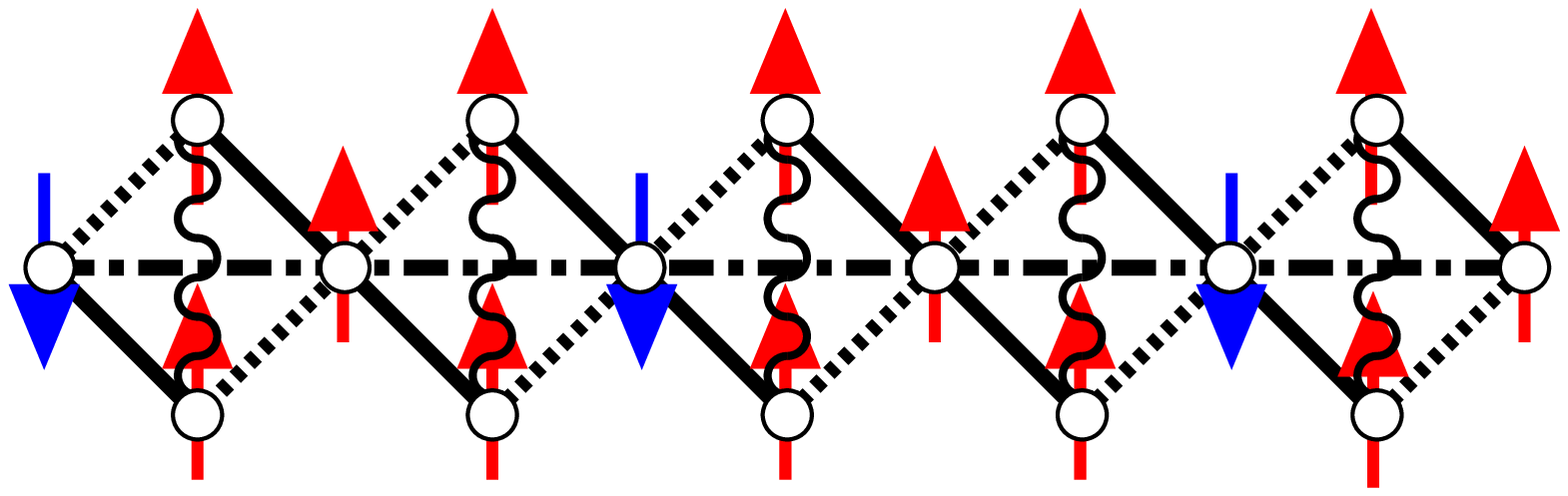}}~~~~~
       }
       \label{fig:pd-m=2-3-config}
       \caption{The spin configurations of the $M=(2/3)\Ms$ plateau states.
       These states are doubly degenerate because the translational invariance is spontaneously broken.
       }
\end{figure}

\section{Discussion}

We have numerically obtained the phase diagrams of the $S=1/2$ distorted diamond type quantum spin chain
with the monomer-monomer interactions and/or the ferromagnetic interactions.
The remarkable nature of the phase diagrams for the $M=\Ms/3$ case is
the existence of the no $M=\Ms/3$ plateau region in the lower-left part.
Since the unit cell of the present model consists of three $S=1/2$ spins,
the appearance of the $M=\Ms/3$ magnetization plateau is naturally expected \cite{oshikawa}.
However, strong quantum fluctuation can destroy this kind of magnetization plateau.
In fact,
in the $S=1/2$ ferromagnetic-ferromagnetic-antiferromagnetic (F-F-AF) trimerized chain \cite{hida},
the $M=\Ms/3$ magnetization plateau vanishes for $|J_{\rm F}/J_{\rm AF}| > 15.4$ \cite{kitazawa-okamoto},
where $J_{\rm F}$ and $J_{\rm AF}$ are the ferromagnetic and antiferromagnetic coupling constants,
respectively.
The relation between the present model and the F-F-AF trimerized chain will be discussed elsewhere.

As stated in (d) of Table 1,
the $M=\Ms/3$ plateau does not exist in the magnetization curve of ${\rm K_3Cu_3AlO_2(SO_4)_4}$.
It seems to be difficult to explain this fact from Fig.3,
because the no $M=\Ms/3$ plateau region appears when $J_3$ is strongly ferromagnetic.

The effect of $J_3$ on the $M=(2/3)\Ms$ magnetization plateau is drastic,
as can be seen from Fig.4.
It is easy to see that the antiferromagnetic $\Jm$ stabilizes the $M=(2/3)\Ms$ magnetization plateau
from the spin configuration of Fig.5.
In fact,
the first order perturbation theory with respect to $\tilde J_2,\tilde J_3$ and $\tilde \Jm$
around the point $\tilde J_2 = \tilde J_3 = \tilde \Jm$ leads to the plateau condition
\begin{equation}
    {5 \over 32}\tilde J_2 + {1 \over 4}\tilde \Jm
    < \tilde J_3
    < {7 \over 16}\tilde J_2 + {5 \over 2}\tilde \Jm
\end{equation}
where the lhs and rhs give the lower right and upper left boundaries, respectively.
The coefficient of $\tilde \Jm$ in the rhs is ten times as large as that in the lhs,
which semi-quantitatively explains the fact that the plateau region strongly extends towards the
upper left direction as $\tilde \Jm$ increases.

We have numerically calculated the magnetization curve by the ED at several  points on the phase diagrams.
All of their behaviors are consistent with the phase diagrams obtained by the LS method.
The details of the magnetization process will be published elsewhere.

\ack

We thank Hikomitsu Kikuchi and Masayoshi Fujihala for informing us their experimental results
prior to publications.
For the  numerical diagonalization, we used the package program TITPACK
ver.2 developed by Hidetoshi Nishimori, to whom we are grateful.
For computer facilities,
we also thank the Supercomputer Center, Institute of Solid State Physics,
University of Tokyo,
and the Computer Room, Yukawa Institute of Theoretical Physics, Kyoto University.


\section*{References}
\begin{thebibliography}{10}

\bibitem{okamoto1}
Okamoto K, Tonegawa T, Takahashi Y and Kaburagi M 1999
{\it J. Phys.: Cond. Mat.} {\bf 11} 10485

\bibitem{tonegawa1}
Tonegawa T, Okamoto K, Hikihara T, Takahashi T and Kaburagi M 2000
{\it J. Phys. Soc. Jpn.} {\bf 69} Suppl. A 332

\bibitem{okamoto2}
Okamoto K, Tonegawa T, Takahashi Y and Kaburagi M 2003
{\it J. Phys.: Cond. Mat.} {\bf 15} 5979

\bibitem{honecker1}
Honecker A and L\"auchli A 2001
{\it Phys. Rev.} B {\bf 63} 174407

\bibitem{kikuchi1}
Kikuchi H, Fujii Y, Chiba M, Mitsudo S, Idehara T, Tonegawa T, Okamoto K,
Sakai T, Kuwai T and Ohta H 2005
{\it Phys. Rev. Lett.} {\bf  94} 227201

\bibitem{kikuchi2}
Kikuchi H, Fujii Y, Chiba M, Mitsudo S, Idehara T, Tonegawa T, Okamoto K,
Sakai T, Kuwai T, Kindo K, Matsuo A, Higemoto W, Nishiyama K, Horovi\'c M
and Bertheir C 2005
{\it Prog. Theor. Phys.} Suppl. No.159 1

\bibitem{gu1}
Gu B and Su G 2006
{\it Phys. Rev. Lett.}  {\bf  97} 089701 

\bibitem{gu2}
Gu B and Su G 2007
{\it Phys. Rev.} B {\bf 75} 174437

\bibitem{kikuchi3}
Kikuchi H, Fujii Y, Chiba M, Mitsudo S, Idehara T, Tonegawa T, Okamoto K,
Sakai T, Kuwai T and Ohta H 2006
{\it Phys. Rev. Lett.} {\bf  97} 089702

\bibitem{mikeska}
Mikeska H-J and Luckmann C 2008
{\it Phys. Rev.} B {\bf 77} 054405 

\bibitem{kang}
Kang J, Lee C, Kremer R K and Whangbo M-H 2009
{\it J. Phys.: Cond. Mat.} {\bf 21} 392201

\bibitem{li}
Li Y-C 2007
{\it J. Appl. Phys.} {\bf 102} 113907

\bibitem{rule}
Rule K C, Wolter A U B,  Su S\"ullow, Tennant D A, Br\"uhl A, K\"ohler S,  Wolf B, Lang M and Schreuer J 2008
{\it Phys. Rev. Lett.} {\bf 100} 1172202

\bibitem{jeschke}
Jeschke H, Opahle I, Kandpal H, Valent\'i R,  Das H, Saha-Dasgupta T, Janson O,
Rosner H, Br\"uhl A,  Wolf B,  Lang M, Richter J, Hu S, Wang X,
Peters R, Pruschke T and Honecker A 2011
{\it Phys. Rev. Lett.} {\bf 106} 217201

\bibitem{honecker2}
Honecker A, Hu S, Peters R and Richter J 2011
{\it J. Phys.: Cond. Mat.} {\bf 23} 164211

\bibitem{kikuchi4}
Kikuchi H, Fujii Y, Matsuo A,  Kindo K 2013
JPS 68th Annual Meeting 26aPS-87
and private communications

\bibitem{kikuchi5}
Asano Y, Kikuchi H, Fujii Y, Matsuo A and Kindo K 2013
JPS 2013 Autumn Meeting 28aKF-6
and private communications
 
\bibitem{yoneyama}
Yoneyama S, Kodama T, Kikuchi K, Fujii Y 2014
Kikuchi H and Fujita W 2014
CrystEngComm {\bf 16} 10385

\bibitem{fujihala}
Fujihala M, Koorikawa H, Mitsuda S, Hagihala M,
Morodomi H, Kawae T, Matsuo A, and Kindo K 2015
{\it J. Phys. Soc. Jpn.} {\bf 84} 073702

\bibitem{koorikawa}
Koorikawa H, Fujihala M, Mitsuda S, Sagayama H, Kumai R, Murakami Y,Nakamura D and Takeyama S 2015
JPS 2015 Autumn Meeting 16aPS-46
and private communications

\bibitem{asano}
Asano T, Ichimura S, Inagaki Y, Kawae T, Matsuo, Kindo K,
Matsumoto M and Kikuchi H 2015
JPS 2015 Autumn Meeting 16aPS-38
and private communications

\bibitem{fujita}
Fujita W, Fujii Y, Kikuchi H 2015
JPS 2015 Autumn Meeting 16aPS-39
and private communications

\bibitem{ls1}
Okamoto K and Nomura K 1992
{\it Phys. Lett.} A {\bf 169} 433

\bibitem{ls2}
Nomura K and Okamoto K 1994
{\it J. Phys. A: Math. Gen.} {\bf 27} 5773

\bibitem{ls3}
Kitazawa A 1997
{\it J. Phys. A: Math. Gen.} {\bf 30} L285

\bibitem{ls4}
Nomura K and Kitazawa A 1998
{\it J. Phys. A: Math. Gen.}, {\bf 31} 7341

\bibitem{oshikawa}
Oshikawa M, Yamanaka M and Affleck I 1997
{\it Phys. Rev. Lett.} {\bf 78} 1984

\bibitem{hida}
Hida K 1994
{\it J. Phys. Soc. Jpn.} {\bf 63} 2359

\bibitem{kitazawa-okamoto} 
Kitazawa A and Okamoto K 1999
{\it J. Phys.: Cond. Mat.} {\bf 11} 9765



\end{thebibliography}
\end{document}